\pdfoutput=1
\newlength{\absize}
\documentclass[12pt]{article}
\usepackage{graphicx,dsfont,subfig,amsmath,amssymb,setspace,sectsty,hyperref}
\hypersetup{linktocpage}
\renewcommand{\baselinestretch}{1.5}

\sectionfont{\large}
\numberwithin{equation}{section}
\begin{document}
\thispagestyle{empty}
\pagestyle{empty}
\renewcommand{\thefootnote}{\fnsymbol{footnote}}
\newcommand{\starttext}{\newpage\normalsize
\pagestyle{plain}
\setlength{\baselineskip}{3.5ex}\par
\setcounter{footnote}{0}
\renewcommand{\thefootnote}{\arabic{footnote}}}
\newcommand{\preprint}[1]{\begin{flushright}
\setlength{\baselineskip}{3ex}#1\end{flushright}}
\renewcommand{\title}[1]{\begin{center}\Large\bf
#1\end{center}\par}
\renewcommand{\author}[1]{\vspace{2ex}{\normalsize\begin{center}
\setlength{\baselineskip}{3.25ex}#1\par\end{center}}}
\renewcommand{\thanks}[1]{\footnote{#1}}
\renewcommand{\abstract}[1]{\vspace{2ex}\normalsize\begin{center}
\centerline{\bf Abstract}\par\vspace{2ex}\parbox{\absize}{#1
\setlength{\baselineskip}{3.25ex}\par}
\end{center}}
\setcounter{bottomnumber}{2}
\setcounter{topnumber}{3}
\setcounter{totalnumber}{4}
\renewcommand{\bottomfraction}{1}
\renewcommand{\topfraction}{1}
\renewcommand{\textfraction}{0}
\def\draft{
\renewcommand{\label}[1]{{\quad[\sf ##1]}}
\renewcommand{\ref}[1]{{[\sf ##1]}}
\renewenvironment{equation}{$$}{$$}
\renewenvironment{thebibliography}{\section*{References}}{}
\renewcommand{\cite}[1]{{\sf[##1]}}
\renewcommand{\bibitem}[1]{\par\noindent{\sf[##1]}}
}
\def\theequation{\thesection.\arabic{equation}}
\preprint{}
\newcommand{\be}{\begin{equation}}
\newcommand{\ee}{\end{equation}}
\newcommand{\ba}{\begin{eqnarray}}
\newcommand{\ea}{\end{eqnarray}}
\newcommand{\bas}{\begin{eqnarray*}}
\newcommand{\eas}{\end{eqnarray*}}
\newcommand{\bc}{\begin{center}}
\newcommand{\ec}{\end{center}}
\newcommand{\nn}{\nonumber}
\newcommand{\comment}[1]{}
\newcommand{\csch}{\mathop{\rm csch\,}}
\newcommand{\sech}{\mathop{\rm sech\,}}

\title{Gravity Dual to Pure $\mathcal{N}=1$ $SU(N)$ Gauge Theory}
\author{Girma Hailu\thanks{hailu@physics.harvard.edu}\\
\vspace{2ex}
\it{Jefferson Physical Laboratory\\
Harvard University\\
Cambridge, MA 02138} }

\abstract{
A correspondence between type IIB string theory with $N$ D7-branes on $\mathds{R}^{1,3}\times \frac{\mathds{C}^1}{Z_2}\times \frac{\mathds{T}^2}{Z_2}\times \frac{\mathds{T}^2}{Z_2}$ and pure $\mathcal{N}=1$ $SU(N)$ gauge theory in four dimensions is proposed and argued. First the supergravity background of unwrapped and flat D7-branes with running axion and dilaton on $\mathds{R}^{1,7}\times \frac{\mathds{C}^1}{Z_2}$ is studied together with the corresponding $\mathcal{N}=1$ $SU(N)$ gauge theory in eight dimensions. The D7-branes are then wrapped over a 4-cycle on  $\frac{\mathds{T}^2}{Z_2}\times \frac{\mathds{T}^2}{Z_2}$ which turns on all $F_1$, $F_3$, $H_3$, and $F_5$ fluxes of type IIB theory and induces torsion. The supergravity solutions are explicitly constructed with exact analytic expressions for all components of the metric and the fluxes. The background geometry of the four-dimensional gauge theory is compact and conformally Calabi-Yau. The internal space normal to the wrapped D7-branes at the infrared boundary is $\mathds{S}^1$ whose radius is set by the nonperturbative scale of the gauge theory and spacetime is $\mathds{R}^{1,3}$ at the ultraviolet boundary. The gauge coupling of the four-dimensional gauge theory is related to the gauge coupling of the eight-dimensional gauge theory and the volume of the 4-cycle. The gravity theory reproduces the renormalization group flow and the pattern of chiral symmetry breaking of the gauge theory and leads to confinement. The curvature is small and nearly constant and the supergravity flow is smooth in the infrared region where the gauge theory is strongly coupled and a dual gravity description is useful. String loop corrections are small for large $N$. The scale of string tension in four dimensions is of the same order as the scale of Kaluza-Klein masses.
}

\starttext
\newpage
\tableofcontents

\setcounter{equation}{0}
\section{\label{sec:intro}Introduction}

There has been interest in constructing a gravity dual to pure $\mathcal{N}=1$ supersymmetric $SU(N)$ gauge theory since the finding of the AdS/CFT correspondence \cite{Maldacena:1998re,Gubser:1998bc,Witten:1998qj}. See \cite{Klebanov:2000hb} and  \cite{Maldacena:2000yy} for two foundational works.
A gravity dual to pure $\mathcal{N}=1$ $SU(N)$ gauge theory is desirable for several reasons. $\mathcal{N}=1$ $SU(N)$ gauge theory exhibits gauge coupling running, confinement, and chiral supersymmetry breaking  and could serve as a laboratory to gain new insight into the theory of quantum chromodynamics (QCD) of the strong nuclear interactions. A crucial feature of QCD is asymptotic freedom \cite{Gross:1973id,Politzer:1973fx} whereby the renormalization group flow of the gauge coupling gives rise to confinement and chiral symmetry breaking. QCD becomes highly nonperturbative at low-energies.
The gauge/gravity duality relates a gauge theory in strongly coupled nonperturbative region to a gravity theory in weakly coupled perturbative region, and it provides the possibility for a calculable classical gravity description to low-energy QCD.
Furthermore, there is a possibility that $\mathcal{N}=1$ supersymmetry may be part of nature at energies accessible in the current generation of experiments at the Large Hadron Collider and a gravity description would be useful for studying nonperturbative phenomena such as mass spectra in the gauge theory.
The first observation that QCD might in some limit be dual to a worldsheet  theory of strings was made in \cite{'tHooft:1973jz}. An additional ingredient which facilitated the development of the gauge/gravity duality was understanding the role of D-branes in string theory \cite{Polchinski:1995mt}.

The first concrete example of gauge/gravity duality was found in \cite{Maldacena:1998re} and it related $\mathcal{N}=4$ $SU(N)$ conformal field theory that lives on a stack of $N$ parallel D3-branes to type IIB string theory on $AdS_5\times \mathds{S}^5$. In \cite{Klebanov:2000hb}, type IIB string theory with $N$ D3- and $M$ D5-branes on $AdS_5\times T^{1,1}$ was found to correspond to $\mathcal{N}=1$ $SU(N+M)\times SU(N)$ gauge theory with bifundamental matter fields and a quartic superpotential with the gravity theory near the infrared (IR) boundary corresponding to $SU(2M)\times SU(M)$ gauge theory with meson and baryon fields \cite{Gubser:2004qj, Hailu:2006uj} rather than to pure confining $SU(M)$ gauge theory as it was originally hoped. In \cite{Maldacena:2000yy}, type IIB string theory with $N$ D5-branes on $\mathds{R}^{1,3}\times \mathds{R}^1\times \mathds{S}^2\times \mathds{S}^3$ was found to reproduce the pattern of chiral symmetry breaking of pure $\mathcal{N}=1$ $SU(N)$ gauge theory but not its renormalization group flow. In both \cite{Klebanov:2000hb} and \cite{Maldacena:2000yy}, the geometry is smooth but noncompact with the size of the internal space increasing in the ultraviolet (UV) and the scale of string tension in four dimensions (4D) is much bigger than the scale of Kaluza-Klein (KK) and glueball masses.

A correspondence between type IIB string theory with $N$ D7-branes on $\mathds{R}^{1,3}\times \frac{\mathds{C}^1}{Z_2}\times \frac{\mathds{T}^2}{Z_2}\times \frac{\mathds{T}^2}{Z_2}$ and pure $\mathcal{N}=1$ $SU(N)$ gauge theory in 4D is proposed and argued in this paper.
The gravity theory reproduces the renormalization group flow of the gauge theory, matches the pattern of chiral symmetry breaking, and leads to confinement.
The background geometry is compact and the internal space normal to the D7-branes is  $\mathds{S}^1$ at the IR boundary and spacetime is $\mathds{R}^{1,3}$ at the UV boundary, consistent with the symmetries of the gauge theory and the radius of $\mathds{S}^1$ is set by the nonperturbative scale of the gauge theory. The supergravity flow is smooth and the curvature of the compact space is smallest in the IR region where the gauge theory is strongly coupled and a dual gravity description is useful.
The 4D UV boundary provides a setting for putting quarks and antiquarks and could also serve as an ultraviolet cutoff to the gravity theory beyond which perturbative gauge theory description is appropriate.
String loop corrections are small for large $N$. The classical supergravity approximation of the string theory accommodates a range of physically interesting values of 't Hooft coupling; the magnitude of the 't Hooft coupling varies from its largest value in the IR to its smallest value in the UV. The scale of string tension in 4D is of the same order as KK and glueball masses, which is useful for exploring mass spectra of glueballs and hadrons. The D7-branes do not disappear and serve as sources of color charge where open strings could end and quarks and antiquarks get bound into hadrons in the supergravity background. The compact extra dimensional space also allows to obtain finite Newton's gravitational constant in 4D and the geometry does not need to be glued to a larger background.

The organization of the paper is as follows. A brief summary of basic features of pure $\mathcal{N}=1$  $SU(N)$ gauge theory in 4D that are relevant to our discussion is given before moving to the construction of the gravity theory.

We note that D7-branes provide a suitable setting which accommodates the symmetries of the gauge theory. In particular, D7-branes in type IIB theory have codimension two which we organize such that one of these dimensions is a radial coordinate that will be mapped to the scale of the gauge theory and the other is angular coordinate that will be mapped to the Yang-Mills angle in the gauge theory. Therefore, we focus on D7-branes and the $F_1$ flux they source as a crucial starting point and seek a background with appropriate 4-cycle for wrapping the D7-branes.

The eight-dimensional (8D) $\mathcal{N}=1$ $SU(N)$ gauge theory that lives on unwrapped and flat $N$ D7-branes on $\mathds{R}^{1,7}\times \frac{\mathds{C}^1}{Z_2}$ background is studied first. The unwrapped D7-branes source the axion potential and induce a running dilaton. The internal space normal to the D7-branes is warped with singularity in the infrared and the background of the 8D gauge theory is noncompact.

The $\mathds{R}^{1,3}\times \frac{\mathds{C}^1}{Z_2}\times \frac{\mathds{T}^2}{Z_2}\times \frac{\mathds{T}^2}{Z_2}$ geometry supports $\mathcal{N}=1$ supersymmetry in 4D and a metric ansatz is written down.
The stack of D7-branes are wrapped over a 4-cycle on $\frac{\mathds{T}^2}{Z_2}\times\frac{\mathds{T}^2}{Z_2}$ at the singularity point of the background geometry of the 8D gauge theory while filling flat 4D spacetime $\mathds{R}^{1,3}$. The D7-branes source the axion potential and the corresponding $F_1$ flux.
The backreaction of the background geometry to the $F_1$ flux induces a $B_2$ potential, and thereby $H_3$ flux. The wrapped D7-branes filling each $\frac{\mathds{T}^2}{Z_2}$ inside $\frac{\mathds{T}^2}{Z_2}\times \frac{\mathds{T}^2}{Z_2}$ are fractional D5-branes which, with $F_1$ and $H_3$, induce $F_3$ flux.
The D7-branes wrapping $\frac{\mathds{T}^2}{Z_2}\times \frac{\mathds{T}^2}{Z_2}$ are also fractional D3-branes which, with $H_3$ and $F_3$, induce $F_5$ flux. Thus the D7-branes provide a setting in which all fluxes of type IIB theory are induced.

With the fluxes turned on, the background backreacts and develops torsion. The background has $SU(3)$ structures and $\mathcal{N}=1$ supersymmetry in 4D is preserved by a balance between the flux and the torsion components.
The metric and all components of the fluxes are explicitly constructed using the equations of motion we wrote in
\cite{Hailu:2007ae} which facilitate systematically studying type IIB flows with $\mathcal{N}=1$ supersymmetry together with the bosonic supergravity equations of motion.  The equations in \cite{Hailu:2007ae} were generalizations of those obtained in \cite{Grana:2004bg} and \cite{Butti:2004pk} using $SU(3)$ structures. The supergravity flow is smooth in the IR and the extra dimensional space is compact and conformally Calabi-Yau with singularity at the UV boundary. The dilaton turns out to be constant now in the background of the 4D gauge theory. The orbifold singularities on $\frac{\mathds{C}^1}{Z_2}\times \frac{\mathds{T}^2}{Z_2}\times \frac{\mathds{T}^2}{Z_2}$ are smoothed out by fluxes. The metric does not involve $AdS_5$ space, since it is related to a gauge theory that is far from conformal.

It is then shown that the solutions to the supersymmetry equations of motion with the Bianchi identities imposed on them solve each and every one of all the bosonic supergravity equations of motion.

The gauge coupling of the 4D gauge theory with the D7-branes wrapped over the 4-cycle is obtained in terms of the gauge coupling of the 8D gauge theory on the unwrapped D7-branes and the volume of the 4-cycle. The gauge coupling of the 4D gauge coupling runs by inheriting the running dilaton from the background of the 8D gauge theory of unwrapped D7-branes. We exploit the simplicity of the two directions normal to the D7-branes in 10D to map one to the scale and the other to the Yang-Mills angle of the 4D gauge theory. The gravity theory reproduces the exact  renormalization group flow of the 4D gauge theory. The range of the radial direction on $\mathds{R}^1$ in $\frac{\mathds{C}^1}{Z_2}\sim \mathds{R}^1\times \mathds{S}^1$ gets smaller for larger 't Hooft coupling at the UV boundary, consistent with the strength of gauge coupling running in the gauge theory.

The magnitudes of the string and the 't Hooft couplings and the curvature of the internal space are then analyzed. One common issue in previous examples of gauge/gravity construction is that the string tension measured by a 4D observer increases while KK and glueball masses decrease with increasing 't Hooft coupling. Consequently, the scale of KK masses is much smaller than the scale of string tension for large 't Hooft coupling. The scale of KK masses and the scale of string tension are of the same order in our construction. The scale of string tension and the size of the transverse space at the IR boundary in the gravity theory are set by the nonperturbative scale of the gauge theory.

The axion potential breaks the $U(1)$ symmetry to $Z_{2N}$ which is further broken down to $Z_2$ by the symmetries of the background giving rise to $N$ discrete vacua in the IR, matching the pattern of chiral symmetry breaking in the gauge theory.

\section{Gauge theory}

Let us make a brief summary of some basic features of pure $\mathcal{N}=1$  $SU(N)$ gauge theory in 4D that are relevant to our discussion. The matter content is a gauge boson and its fermionic superpartner, a gaugino, both transforming in the adjoint representation.
The gauge theory has a coupling coefficient $\tau=\frac{4\pi i}{g_4^2}+\frac{\Theta}{2\pi}$, where $g_4$ is the Yang-Mills coupling and $\Theta$ is the Yang-Mills angle.

At the classical level, the theory has global $U(1)$ R-symmetry under which the gaugino field $\lambda^\alpha$ transforms as $\lambda^\alpha\to e^{i\,c}\lambda^{\alpha}$, $c\in \mathds{R}$ which is equivalent to shifting the Yang-Mills angle $\Theta\to \Theta+2Nc$ and
the $U(1)$ symmetry is anomalous in the quantum theory.
Because $\Theta\sim\Theta+2\pi k$, the quantum theory has a reduced anomaly-free discrete $Z_{2N}$ symmetry.
Gaugino condensation, which results in $\langle\mathrm{tr}\, \lambda^\alpha \lambda_\alpha\rangle\ne0$, breaks the $Z_{2N}$ symmetry down to $Z_2$ giving $N$ number of discrete vacua.

The low-energy IR dynamic of the theory at the scale $\Lambda$ is described by the Veneziano-Yankielowicz
superpotential \cite{Veneziano:1982ah},
\begin{equation}
W_{\mathrm{VY}}=NS-NS\log(\frac{S}{\Lambda^{3}}),\quad S=-\frac{1}{32\pi^2}\mathrm{Tr\,}\mathcal{W}^{\alpha}
\mathcal{W}_{\alpha},\label{eq:rev4-5}
\end{equation}
where $S$ is the glueball superfield defined in terms of the gauge
chiral superfield $\mathcal{W}_{\alpha}$ containing the gauge and the gaugino fields. Extremizing the superpotential $W_{\mathrm{VY}}$ with $S$ gives the vacuum expectation values of the glueball superfield corresponding to the $N$ vacua,
\be
\langle S \rangle= \Lambda^{3} e^{\frac{2\pi i k}{N}},\quad  k=1, 2,\cdots,N.
\ee

\begin{figure}[t]
\begin{center}
\includegraphics[width=3.25in]{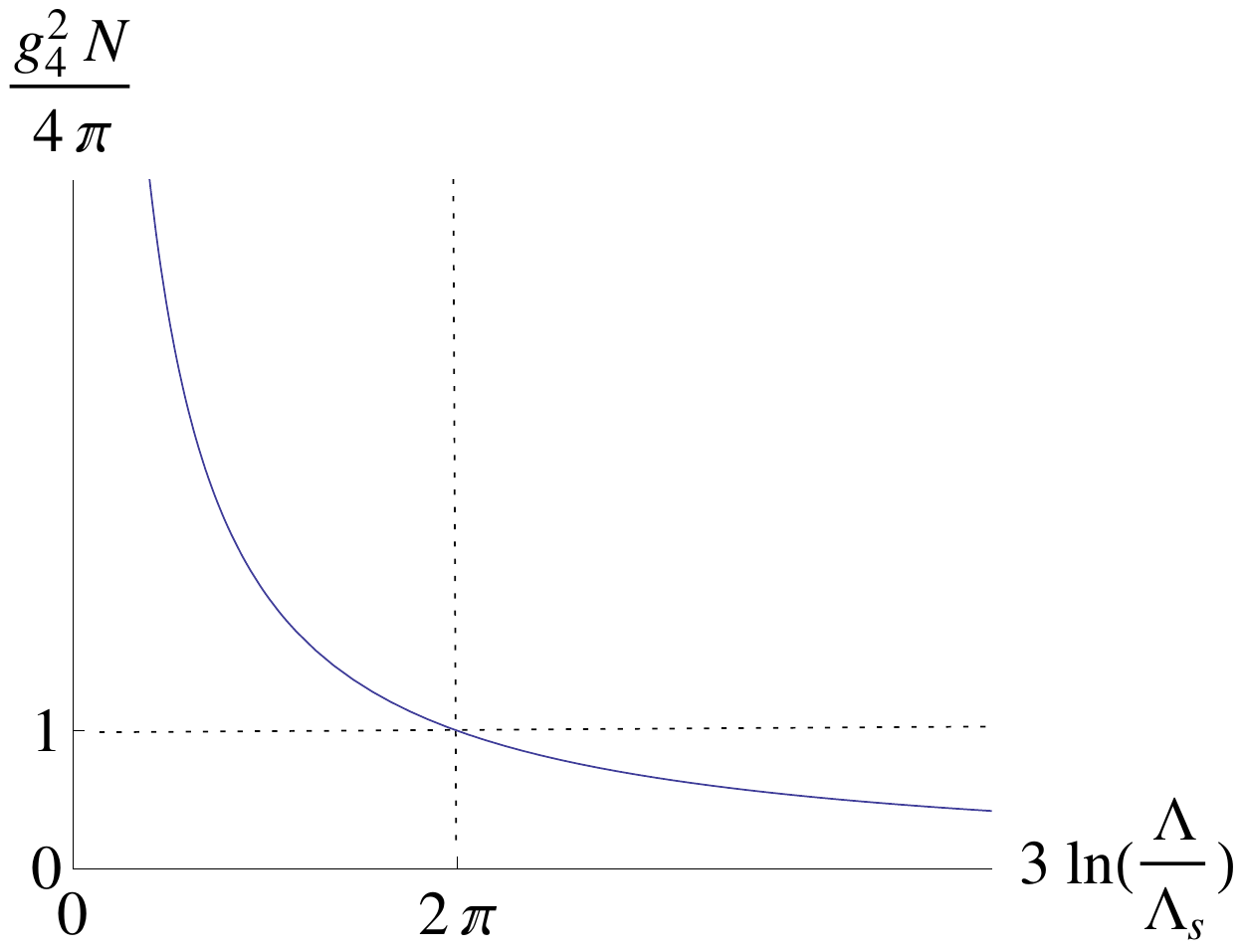}
\end{center}
\caption{Gauge coupling running.}
\label{fig:g4coup}
\end{figure}

Quantum corrections to the renormalization group flow of the holomorphic coupling coefficient are exhausted at one loop and with the exact $\beta$ function we write
\be \frac{4\pi}{g_4^2}=\frac{3N}{2\pi}\ln(\frac{\Lambda}{\Lambda_s}),\quad \label{dTdmu-1}
\ee
where we take $\Lambda_s$ to be the nonperturbative scale of the theory at which the gauge coupling formally diverges. The running of the gauge coupling is shown in Figure \ref{fig:g4coup}. The value of $\frac{g_4^2N}{4\pi}=1$ at $3\,\ln(\frac{\Lambda}{\Lambda_s})=2\pi$. For $3\,\ln(\frac{\Lambda}{\Lambda_s})> 2\pi$ the gauge theory has perturbative description. Our interest is in a gravity description in the region where $\frac{g_4^2N}{4\pi}>>1$ and the gauge theory is highly nonperturbative. We call the corresponding range of energy scales where the gauge theory is highly nonperturbative low-energy.

\section{Unwrapped D7-branes\label{sec-d7-dyn}}

Before we start wrapping the D7-branes to construct a 4D gauge theory, we need to look at the 8D gauge theory that lives on unwrapped and flat $N$ D7-branes.
Consider a stack of $N$ parallel D7-branes which fill flat 8D spacetime $\mathds{R}^{1,7}$ on $\mathds{R}^{1,7}\times \frac{\mathds{C}^1}{Z_2}$ background.
Preserving $\mathcal{N}=1$ supersymmetry in the 8D theory requires that the internal  two-dimensional space be complex.
Noting that $\frac{\mathds{C}^1}{Z_2}\sim \mathds{R}^1\times \mathds{S}^1$, the complex coordinate on $\mathds{C}^1$ is defined in terms of a radial variable $r$ on $\mathds{R}^1$ and an angular coordinate $\psi$ on $\mathds{S}^1$ as $z^1=(\frac{r}{r_s})^3\, e^{i\psi}=e^{\rho+i\,\psi}$,
where
\be
\rho\equiv 3\ln(\frac{r}{r_s})\label{tbar-1}
\ee
and $r_s$ is a constant with dimension of length.
The coordinate on $\mathds{S}^1$ is parameterized by $\psi\in[0,2\pi]$.
The $Z_2$ symmetry transforms $z^1\to -z^1$ and imposes the relation $\psi\sim \psi+\pi$ and the orbifold has one fixed point at $r=0$. The D7-branes are located at the singularity.
We have introduced $\rho$ and the dimensionful parameter $r_s$ because, as we see in the supergravity solutions below, there is a logarithmic flow along $\mathds{R}^1$ with $\mathds{S}^1$ warped to zero-size at $r=r_s$ and the radial supergravity flow is constrained to $r\ge r_s$ or $\rho\ge0$.  The physical interpretation of $r_s$ will be discussed later in this and the following sections.

Because the dilaton runs in the background of the 8D gauge theory, we find the Einstein frame, in which the gauge coupling is dilaton dependent while the gravitational constant is fixed, convenient. We will use the string frame when studying the background of the 4D gauge theory with all fluxes turned on in the remaining sections, since the dilaton appears in the same form in all the R-R flux terms in the supersymmetry transformations in the string frame which we find more convenient.
The dilaton will in the end turn out to be constant in the background of the 4D gauge theory and the final expressions in the 4D gauge theory background can be interpreted as quantities in the same Einstein frame as in the 8D gauge theory background.

We take the metric ansatz with flat 8D spacetime as
\be
ds_{10}^2=dx_{1,7}^2+r_s^2\,e^{2H}(d\rho\,^2+d\psi^2),\label{metric-2bb}
\ee
where $H=H(\rho)$ and $dx_{1,7}^2$ is the metric on flat $\mathds{R}^{1,7}$.
The  D7-branes at $r=0$ produce axion potential $C_0$ and the corresponding  $F_1$ flux,
\be
C_0=\frac{N}{2\pi}\psi, \qquad F_1=dC_0=\frac{N}{2\pi}d\psi, \label{F1C0-1}
\ee
where the $F_1$ flux is normalized as $\int_{0}^{2\pi}F_1=N$.

Type IIB theory has two Majorana-Weyl spinors of the same chirality with a total of 32 supercharges. The unwrapped D7-branes reduce the number of preserved supercharges by half.
In order for the background with the flux turned on to preserve $\mathcal{N}=1$ supersymmetry, the axion-dilaton coupling $\tau=C_0+i\,e^{-\Phi}$ needs to be a holomorphic or anti-holomorphic function of the complex coordinate on the normal space $\mathds{R}^1\times \mathds{S}^1$. Thus the $F_1$ flux on the above background metric induces the dilaton.
The supergravity equations in the bosonic sector in Einstein frame reduce to
\ba
&\frac{1}{\sqrt{-G}}\,\partial_M\left(\sqrt{-G}\,
G^{MN}\partial_N\Phi\right)
=e^{2\Phi}F_1^2,&\nn\\
&R_{MN}=\frac{1}{2}\,\partial_M\Phi \partial_N\Phi +\frac{1}{2}e^{2\Phi}\partial_M C_0\partial_N C_0,\label{phi-RMN-1a}&
\ea
and we take the ansatz $\Phi=\Phi(\rho)$ which accommodates a holomorphic or anti-holomorphic axion-dilaton coupling for the $C_0(\psi)$ we have here. The components of the equations given by (\ref{phi-RMN-1a}) are $
\frac{d^2\Phi}{d\rho^2}=e^{2\Phi}\left(\frac{N}{2\pi}\right)^2$ and
$\frac{1}{2}(\frac{d\Phi}{d\rho})^2=-\frac{d^2H}{d\rho^2}$.
The solutions are
\be
e^{-\Phi}=\frac{N}{2\pi}\,\rho,\qquad e^{2H}=\rho,\label{phi-T-8d-1}
\ee
where we have set $e^{-\Phi(0)}=0$.
Notice that $\rho=0$ is a singularity point with $\mathds{S}^1$ warped to zero-size, and the physical region for the radial supergravity flow is $\rho\ge0$. The axion-dilaton coupling is $\tau=\frac{N}{2\pi}(\psi+i\,\rho)$, which is anti-holomorphic in our notation.

We can check that the energy from the curvature of the metric cancels out that from the axion and the dilaton fields in the supergravity action. In particular, $R=(\partial \Phi)^2=e^{2\Phi}F_1^2=r_s^{-2}\rho\,^{-3}$ and the 10D action which is now proportional to
$\int d^{10}x\,\sqrt{-G}(R$ $-\frac{1}{2}(\partial \Phi)^2-\frac{1}{2}e^{2\Phi}F_1^2)$
vanishes.

The Yang-Mills coupling of the 8D gauge theory that lives on the D7-branes is then given by $g_8$ such that
\be
g_8^2=2(2\pi)^5 \alpha'^2 e^\Phi=\frac{2(2\pi)^6\alpha'^2}{N\rho}.\label{g8-1}
\ee
The 't Hooft coupling $g_8^2 N$ is large and the gauge theory is strongly coupled in the IR region near $\rho=0$. The string coupling $e^{\Phi}$ can be made small in all the region $\rho>0$ for appropriately large $N$. The curvature is large in the region near the IR boundary. See section \ref{sec-4dgc} for discussion on $\alpha'$ and string loop corrections.

Thus $r=r_s$ or $\rho=0$ is the location on $\mathds{R}^1$ where the dilaton and the 8D gauge coupling diverge, $\mathds{S}^1$ is warped to zero-size, and the geometry is singular. After the D7-branes are put at the fixed point $r=0$ of $\frac{\mathds{C}^1}{Z_2}$, the physical region of the warped supergravity background on $\mathds{R}^1$ where the metric is positive definite and the dilaton is real  is $r\ge r_s$ or $\rho\ge 0$.

\section{Metric ansatz\label{sec-metricansatz}}

We seek a metric with appropriate 4-cycle inside the extra six-dimensional (6D) space $Y$ of a 10D spacetime of the form $\mathds{R}^{1,3}\times Y$, where $Y$ contains $\frac{\mathds{C}^1}{Z_2}\sim \mathds{R}^1\times \mathds{S}^1$, for wrapping the D7-branes over at the singular point $\rho=0$ on $\mathds{R}^1$ of the background geometry of the 8D gauge theory such that the supersymmetry will be $\mathcal{N}=1$ in 4D. This requires that $Y$ be complex.

Consider the orbifold $Y=\frac{\mathds{C}^1}{Z_2}\times \frac{\mathds{T}^2}{Z_2}\times \frac{\mathds{T}^2}{Z_2}$. The complex coordinate on $\mathds{C}^1$ is denoted by the same variable $z^1$. Let us denote the complex coordinates on the two ${\mathds{T}^2}$'s by $z^2$ and $z^3$. The three $Z_2$ transformations are generated by the operators $g_1:\, (z^1,\,z^2,\,z^3)\to (-z^1,\,z^2,\,z^3)$,  $g_2:\, (z^1,\,z^2,\,z^3)\to (z^1,\,-z^2,\,z^3)$, and $g_3:\, (z^1,\,z^2,\,z^3)\to (z^1,\,z^2,\,-z^3)$. Let us decompose the two 16-component Majorana-Weyl spinor parameters of the same chirality for the supersymmetry transformations of type IIB theory in 10D, denoted by $\epsilon^{1}$ and $\epsilon^{2}$, as
$\epsilon^1=\zeta^1_{+}\otimes\eta_{+}+\zeta^1_{-}\otimes\eta_{-}$ and  $\epsilon^2=\zeta^2_{+}\otimes\eta_{+}+\zeta^2_{-}\otimes\eta_{-}$; $\zeta^1_{\pm}$ and $\zeta^2_{\pm}$ are 2-component positive and negative chirality Weyl spinors on $\mathds{R}^{1,3}$ such that $\zeta^{1,2}_{-}={\zeta^{1,2}_{+}}^*$ and $\eta_{\pm }$ are 4-component positive and negative chirality Weyl spinors on $Y$ such that $\eta_{-}={\eta_{+}}^*$. We choose the three Cartan generators of the Clifford algebra on $Y$ to be ones which correspond to rotations on each factor of $Y$ with complex coordinates $z^1$, $z^2$, and $z^3$. The eigenvalues of the Cartan generators are   $(\pm\frac{1}{2},\,\pm\frac{1}{2}\,\pm\frac{1}{2})$. Only the spinor ${\eta_{+\frac{1}{2},+\frac{1}{2},+\frac{1}{2}}}$ is invariant under ${Z_2\times Z_2\times Z_2}$ and the $\mathds{R}^{1,3}\times\frac{\mathds{C}^1}{Z_2}\times \frac{\mathds{T}^2}{Z_2}\times \frac{\mathds{T}^2}{Z_2}$ background supports $\mathcal{N}=1$ supersymmetry in 4D.

The radial coordinate on $\mathds{R}^1$ in $\frac{\mathds{C}^1}{Z_2}\sim \mathds{R}^1\times \mathds{S}^1$ is now parameterized by a dimensionless variable $t$ such that $t=0$ corresponds to $\rho=0$ or $r=r_s$. The range of $t$ and its relation to $\rho$ or $r$ will be determined later when the supergravity solutions are established. The angular coordinate on $\mathds{S}^1$ is denoted by the same variable $\psi$. The two angular coordinates on one of the $\mathds{T}^2$'s are denoted by $\varphi_1$ and $\varphi_2$ and on the other by $\varphi_3$ and $\varphi_4$. The ansatz for the 10D metric is
\be
ds_{10}^2= e^{2A}dx_{1,3}^2+r_s^2\,e^{2T}dt^2+r_s^2\,e^{2P}d\psi^2
+r_s^2\,e^{2Q}\left(d\varphi_1^2+d\varphi_2^2+d\varphi_3^2+d\varphi_4^2 \right).\label{metric-1a}
\ee
The 4D spacetime $\mathds{R}^{1,3}$ is flat with coordinates denoted by $x^\mu$, $\mu=0,1,2,3$, and the metric denoted by $dx_{1,3}^2$. The angular coordinates $\psi$, $\varphi_1$, $\varphi_2$, $\varphi_3$, and $\varphi_4$ are all defined to have the same range $[0,2\pi]$. The variables $A$, $T$, $P$, and $Q$ are all functions of $t$.
The D7-branes will be wrapped over $\frac{\mathds{T}^2}{Z_2}\times \frac{\mathds{T}^2}{Z_2}$ at $r=r_s$, the singularity point of the background of the 8D gauge theory. The value of $\psi$ on $\mathds{S}^1$ where the D7-branes are wrapped is related to chiral symmetry breaking which will be discussed in section \ref{sec-csb}.

We introduce the following complex basis with holomorphic/anti-holomorphic indices which we use in studying the supersymmetry constraints,
\ba
Z^1&= &r_s\,\left(e^{T}\,dt+i\,e^{P}\,d\psi\right), \quad \bar{Z}^{\bar{1}}= r_s\,\left(e^{T}\,dt-i\,e^{P}\,d\psi\right),\quad\nn\\  Z^2&= &r_s\,e^{Q}\left(d\varphi_1+i\,d\varphi_2\right), \quad
\bar{Z}^{\bar{2}}= r_s\,e^{Q}\left(d\varphi_1-i\,d\varphi_2\right),
\quad\nn\\ Z^3&= &r_s\,e^{Q}\left(d\varphi_3+i\,d\varphi_4\right),\quad \bar{Z}^{\bar{3}}= r_s\,e^{Q}\left(d\varphi_3-i\,d\varphi_4\right).\quad\label{hol-anihol-1}
\ea
The complex and the K\"ahler structures are described in terms of the following  K\"ahler 2-form $J$ and holomorphic 3-form $\Omega$,
\be
J=
\frac{i}{2} \delta_{j\bar{k}}{Z^{j}}{ {\wedge} }{\bar{Z}^{\bar{k}}},\qquad
\Omega =
{Z^1}{ {\wedge} }{Z^2}{ {\wedge} }{Z^3}.\label{JZ-OmZ-1}
\ee

\section{Supersymmetry equations of motion\label{sec-susyeom}}

Wrapping the D7-branes over $\frac{\mathds{T}^2}{Z_2}\times \frac{\mathds{T}^2}{Z_2}$ at $t=0$ turns on fluxes and the  background geometry backreacts and develops torsion.
In order to preserve $\mathcal{N}=1$ supersymmetry in 4D, the two positive chirality Weyl spinors $\zeta^1_+$ and $\zeta^2_+$ in 4D that we discussed in section \ref{sec-metricansatz} are taken to be proportional to each other.
Because the 4D spacetime is flat, we can write $\zeta_+^1=\frac{1}{2}(\alpha+\beta) \zeta_+$ and
$\zeta_+^2=\frac{1}{2i}(\alpha-\beta) \zeta_+$, where $\alpha$ and $\beta$ are complex parameters which depend only on the coordinates on $Y$.
With these, the spinor decompositions is rewritten as $\epsilon^1=\zeta_{+}\otimes\eta^1_{+}+\zeta_{-}\otimes\eta^1_{-}$ and $\epsilon^2=\zeta_{+}\otimes\eta^2_{+}+\zeta_{-}\otimes\eta^2_{-}$, where
$\eta_+^1=\frac{1}{2}(\alpha+\beta) \eta_+$ and
$\eta_+^2=\frac{1}{2i}(\alpha-\beta) \eta_+$.
We use $SU(3)$ structures such that $\eta_{+}\equiv {\eta_{+\frac{1}{2},+\frac{1}{2},+\frac{1}{2}}}$
is the globally invariant $SU(3)$ singlet spinor on $Y$. The constraint leads to a set of relations among the components of the torsion, the fluxes, and the spinor parameters organized in $SU(3)$ representations.
See \cite{Hailu:2007ae, Grana:2004bg, Butti:2004pk} for details and for more on our notation.

The metric on the 10D spacetime is written as $ds_{10}^2=e^{2A}dx_{1,3}^2 +ds_6^2$, where $A$ here is generally a function of the coordinates on the extra 6D space and the metric on the extra space is expressed in terms of complex 1-forms with holomorphic/antiholomorphic indices as $ds_6^2=\delta_{i\bar{j}}Z^{i}\bar{Z}^{\bar{j}}$.
The torsion components come in the variations of the fundamental 2-form, $dJ= -\frac{3}{2}\mathrm{Im}$ $(W_1^{(1)} \bar{\Omega})+(W_4^{(3)}+W_4^{(\bar{3})})\wedge J+(W_3^{(6)}+W_3^{(\bar{6})})$ and the holomorphic 3-form,
$d\Omega=W_1^{(1)} J^2+W_2^{(8)}\wedge J+W_5^{(\bar{3})}\wedge \Omega$.  The superscripts denote the $SU(3)$ representations.
The string coupling is written in terms of the dilaton as $e^{\Phi}$.
The components of the 3-form fluxes come in $H_3=-\frac{3}{2}\mathrm{Im}(H_3^{(1)} \bar{\Omega})+(H_3^{(3)}+H_3^{(\bar{3})})\wedge J+(H_3^{(6)}+H_3^{(\bar{6})})$ and
$F_3=-\frac{3}{2}\mathrm{Im}(F_3^{(1)} \bar{\Omega})+(F_3^{(3)}+F_3^{(\bar{3})})\wedge J+(F_3^{(6)}+H_3^{(\bar{6})})$.
The self-dual 5-form flux is written as $\tilde{F}_5=(1+\star_{10})F_5$ with
${F_5}=(F_5^{(3)}+F_5^{(\bar{3})})\wedge J\wedge J$.
The 1-form flux is decomposed as $F_1=F_1^{(\bar{3})}+F_1^{(3)}$.

The complete set of equations which describe the balance among the components of the fluxes, the dilaton, the warp factor and the torsion in terms of the complex parameters $\alpha$ and $\beta$ such that $\mathcal{N}=1$ supersymmetry is preserved in 4D are summarized below.
\ba
& W^{(1)}_1=0,\quad W^{(8)}_2=0,\quad F^{(1)}_3=0,\quad H^{(1)}_3=0,\label{1p8-1}&\nn\\
&(\alpha^2-\beta^2)W_3^{(6)}=2\alpha \beta e^{\Phi} F_3^{(6)},
\quad(\alpha^2+\beta^2)W_3^{(6)}=-2\alpha \beta \star_{6} H_3^{(6)},\label{66bar}&\nn\\
& e^{\Phi}\Bigl(F_3^{(\bar{3})}+\frac{2i\alpha \beta}{\alpha^2+\beta^2}F_1^{(\bar{3})}\Bigr)=-\frac{2\alpha \beta}{\alpha^2+\beta^2}\bar{\partial}\ln(\frac{\beta}{\alpha}),\quad e^{\Phi}\Bigl(F_5^{(\bar{3})}+\frac{1}{2}F_1^{(\bar{3})}\Bigr)
=i\bar{\partial}\ln(\frac{\beta}{\alpha}),\quad &\label{F5F33b}\nn\\
&H_3^{(\bar{3})}=-\frac{2i\alpha \beta}{\alpha^2-\beta^2}\bar{\partial}\ln(\frac{\beta}{\alpha}),\quad \bar{\partial}\Phi+i\frac{\alpha^2- \beta^2}{\alpha^2+\beta^2}e^{\Phi}F_1^{(\bar{3})}=\frac{4\alpha^2 \beta^2}{\alpha^4-\beta^4}\bar{\partial}\ln(\frac{\beta}{\alpha}),&
\label{H3phi3b}\nn\\
&\bar{\partial}A+\frac{i}{2}\frac{\alpha^2- \beta^2}{\alpha^2+\beta^2}e^{\Phi}F_1^{(\bar{3})}=-\frac{\alpha^2- \beta^2}{2(\alpha^2+\beta^2)}\bar{\partial}\ln(\frac{\beta}{\alpha}),
\quad
A=\ln(|\alpha|^2+|\beta|^2),&\label{A33bA}\nn\\
&W_4^{(\bar{3})}=\frac{\alpha^2+ \beta^2}{\alpha^2-\beta^2}\bar{\partial}\ln(\frac{\beta}{\alpha}),\quad W_5^{(\bar{3})}=-\frac{3\alpha^2 +\beta^2}{\alpha^2-\beta^2}\bar{\partial}\ln \alpha+\frac{\alpha^2 +3\beta^2}{\alpha^2-\beta^2}\bar{\partial}\ln \beta.&\label{W53b}
\ea

\section{Metric and flux components\label{sec-metric-flux}}

Now we start finding the metric and the flux components for the background with wrapped D7-branes using the metric ansatz written in section \ref{sec-metricansatz} and the supersymmetry equations of motion summarized in section \ref{sec-susyeom}.
The torsion components $W_1$, $W_2$, and $W_3$ vanish for the metric given by (\ref{metric-1a}) and
\be
W_4= 2Q'\,dt,\qquad W_5=(2Q'+P')\,dt,\label{W4W5s-1}
\ee
where a prime denotes differentiation with respect to $t$.
For the spinor parameters $\alpha$ and $\beta$ we use the ansatz
\be
\ln\left(\frac{\beta}{\alpha}\right)=u+i\,\frac{\pi}{2},\label{albe-1}
\ee
where $\alpha$, $\beta$, and $u$ are all functions of $t$.

The wrapped D7-branes at $t=0$ produce the same axion potential $C_0$ and $F_1$ given by \ref{F1C0-1}.
with the $F_1$ flux normalized as $\int_{0}^{2\pi}F_1=N$.
The $F_1$ flux on the background metric given by (\ref{metric-1a}) induces a $B_2$ potential and the corresponding $H_3=dB_2$ flux.
Moreover, the D7-branes wrapping $\mathds{T}^2$'s inside $Y$ are fractional D5-branes, which with the $F_1$ and the $H_3$ fluxes, induce $F_3$ flux. The D7-branes wrapping $\frac{\mathds{T}^2}{Z_2}\times \frac{\mathds{T}^2}{Z_2}$ are fractional D3-branes, which with the $F_3$ and the $H_3$ fluxes, induce $F_5$ flux. Thus all $F_1$, $F_3$, $H_3$, and $F_5$ fluxes of type IIB theory are induced.

We then use the equations of motion given by (\ref{W53b}) to find the relations among the components of the fluxes, the metric, the dilaton, and the spinor parameter $u$,
\ba
&F_3=\left(-\frac{N}{2\pi}+e^{P-T-\Phi}\,u'\right)r_s^2 \,e^{2Q}\,\csch u \, \left(d\psi\wedge d\varphi_1\wedge  d\varphi_2 +d\psi\wedge  d\varphi_3\wedge  d\varphi_4\right),&\label{F3sol-1}\nn\\
&H_3=r_s^2 \,e^{2Q}\,\sech u\, u'\left(dt\wedge d\varphi_1\wedge  d\varphi_2+dt\wedge d\varphi_3\wedge  d\varphi_4\right),&\label{H3sol-1}\nn\\
&\tilde{F}_5=\left(-\frac{N}{2\pi}+2\,e^{P-T-\Phi}\,u'\right)r_s^4 \,e^{2Q}\, d\psi\wedge d\varphi_1\wedge  d\varphi_2\wedge  d\varphi_3\wedge  d\varphi_4&\nn\\&+\left(\frac{N}{2\pi}-2\,e^{P-T-\Phi}\,u'\right)\,
e^{4A+T-P} \, d^4x \wedge dt,  &\label{F5sol-1}\nn\\
& \Phi'=-\frac{N}{2\pi}\,e^{T-P+\Phi}\,\coth u+\csch u\,\sech u\, u',& \nn\\
& A'=-\frac{N}{4\pi}\,e^{T-P+\Phi}\,\coth u+\frac{1}{2}\coth u \, u',&\label{phi-soln-2}\nn\\
&  Q'=-\frac{1}{2}\tanh u \,u',& \nn\\
& P'=\frac{N}{4\pi}\,e^{T-P+\Phi}\,\coth u-\frac{1}{2}\coth u \,u'.\qquad &\label{APQ-soln-1}
\ea

Imposing the Bianchi identity
\be dF_3=-F_1\wedge H_3\qquad \text{or}\qquad  d\tilde{F}_5=H_3\wedge F_3,\label{Bianchi-1}\ee we have
\be
u'=\frac{N}{2\pi}\,e^{T-P+\Phi}\, \cosh^2u.\label{u-soln-1}
\ee
With one of the identities in (\ref{Bianchi-1}) imposed, the other is automatically satisfied. Moreover, using (\ref{u-soln-1})) in the $\Phi'$ equation in (\ref{phi-soln-2}) leads to a constant dilaton and we set
\be e^\Phi=g_s.\label{ephi-1}\ee
Using the equation for $Q'$ and putting (\ref{u-soln-1}) in the equations for $P'$ and $Q'$ in (\ref{u-soln-1}), we obtain
\be
e^{2Q}=e^{2P}=e^{-2A}=\sech u.\label{PQA-soln-2}
\ee

The fluxes can similarly be written down in terms of $u$ using (\ref{u-soln-1}) and (\ref{PQA-soln-2}) in (\ref{APQ-soln-1}), and all fluxes are summarized below.
\ba
&F_1=\frac{N}{2\pi}d\psi,&\nn\\ & F_3=\frac{N}{2\pi}\,r_s^2\,\tanh u \, \left(d\psi\wedge d\varphi_1\wedge  d\varphi_2 +d\psi\wedge  d\varphi_3\wedge  d\varphi_4\right),&\nn\\ &H_3=\frac{g_sN}{2\pi}\,r_s^2\,e^{T}\,\sqrt{\cosh u}\left(dt\wedge d\varphi_1\wedge  d\varphi_2+dt\wedge d\varphi_3\wedge  d\varphi_4\right), &\nn\\
&\tilde{F}_5=\frac{N}{2\pi}\,r_s^4\,\left(1+\tanh^2u\right) \, d\psi\wedge d\varphi_1\wedge  d\varphi_2\wedge  d\varphi_3\wedge  d\varphi_4&\nn\\ &-\frac{N}{2\pi}\,e^{T}\,\cosh 2u\, \cosh^2u \, d^4x \wedge dt. \qquad \,&\label{F5sol-1b}
\ea
Notice that the $H_3$ flux is closed, $H_3=dB_2$, and the $B_2$ potential is given by
\be
B_2=r_s^2\,\tanh u\,\left(d\varphi_1\wedge  d\varphi_2+d\varphi_3\wedge  d\varphi_4\right).\label{B2-soln-1}
\ee

Thus all components of the fluxes and the metric are expressed in terms of $u$ and $T$. The only component of the metric that is not yet determined in terms of $u$ is $T$, which multiplies $dt^2$ and determines the definition of the radial coordinate, and we do that in the next section.

\section{Bosonic supergravity equations of motion}

In this section we determine $T$ in term of $u$ and verify that the solutions solve all bosonic supergravity equations.
We work in the string frame. Upper case indices $M,\,N\,\cdots$ represent the coordinates of the 10D spacetime, the Greek letters $\mu,\,\nu\,\cdots$ represent the coordinates of the 4D spacetime, and lower case indices  $m,\,n\,\cdots$ represent the coordinates of the extra 6D space.

The equation from the variation of the type IIB supergravity action, which can be written with the self-duality of the 5-form flux additionally imposed, with respect to the dilaton is
\be
\frac{1}{\sqrt{-G}}\,\partial_M\left(\sqrt{-G}\,e^{-2\Phi}G^{MN}
\partial_N\Phi\right)=
-\frac{1}{8}F_1^2-\frac{1}{96}F_3^2+\frac{1}{96}e^{-2\Phi}H_3^2.
\label{phi-1}
\ee
The equations from the variation of the action with respect to the metric are
\ba R_{MN}=& &-4\,\partial_M\Phi \partial_N\Phi +\frac{1}{2}e^{2\Phi}\partial_M C_0\partial_N C_0+\frac{1}{4}(H_3)_{MOP}(H_3)_{N}^{\hspace{3mm}OP}\nn\\ & &+\frac{1}{4}e^{2\Phi}(F_3)_{MOP} (F_3)_{N}^{\hspace{3mm}OP}+\frac{1}{96}e^{2\Phi}(\tilde{F}_5)_{MOPQR}
(\tilde{F}_5)_{N}
^{\hspace{3mm}OPQR}\nn\\ & &
-G_{MN}\,(\frac{1}{48}H_3^2+\frac{1}{48}e^{2\Phi}F_3^2
+\frac{1}{960}e^{2\Phi}\tilde{F}_5^2),
\label{RMN-1}
\ea
where $G_{MN}$ represent components of the metric and $e^{\Phi}=g_s$ in our case.

Because we know all the metric and the flux components in term of $u$ and $T$, it is straightforward to write the right hand sides of (\ref{phi-1}) and (\ref{RMN-1}) in terms of $u$ and $T$ using (\ref{ephi-1})-(\ref{F5sol-1b}).
Our notation is such that the 10D volume element is
\be
\mathrm{vol}={r_s^6\,e^{4A+T+P+4Q}}\,dx^0\wedge
dx^1\wedge dx^2 \wedge dx^3 \wedge dt \wedge d\psi \wedge d\varphi_1
\wedge d\varphi_2 \wedge d\varphi_3  \wedge d\varphi_4
\label{vol10d}
\ee
and the inner product of a $p$-form
$\omega_p=\frac{1}{p!}(\omega_p)_{M_1 \dots M_p}$ $dx^{M_1}\wedge \dots
\wedge dx^{M_p}$ and its Hodge star in 10D satisfy
\begin{equation}
\omega_p \wedge \star_{10} \,\omega_p= \frac{1}{p!}\, (\omega_p)_{M_1 \dots
M_p} (\omega_p)^{M_1 \dots M_p}\, \mathrm{vol}= \frac{1}{p!}\,
\omega_p^2\, \mathrm{vol}.\label{wpinner}
\end{equation}

The magnitudes of the fluxes are
\ba
&F_1^2=\frac{1}{r_s^2}\left(\frac{N}{2\pi}\right)^2\,\cosh u,\quad F_3^2=\frac{12}{r_s^2}\left(\frac{N}{2\pi}\right)^2\,\sinh^2 u\,\cosh u,,&\nn\\& H_3^2=\frac{12}{r_s^2}\left(\frac{g_sN}{2\pi}\right)^2 \,\cosh^3 u,\quad \tilde{F}_5^2=0.&\label{flux-mag-1}
\ea
The components of the Ricci tensor are
\ba
R_{\mu\nu}&=&\frac{1}{4\,r_s^2}\left(\frac{g_sN}{2\pi}\right)^2\,\cosh 2u\,
\cosh u \left(1-e^{2T}\,\cosh 2u\, \cosh u \right) \,\,G_{\mu\nu},\nn\\
R_{tt}&=&\frac{1}{2\,r_s^2}\left(\frac{g_sN}{2\pi}\right)^2(2-\cosh 2u)\, \cosh^3 u \,G_{tt},\nn\\
R_{m'n'}&=&\frac{1}{2\,r_s^2}\left(\frac{g_sN}{2\pi}\right)^2\cosh 2u\, \cosh^3 u \,G_{m'n'},
\label{RMN-2}
\ea
where $m'$ and $n'$ run over the angular coordinates.

Noticing that $2A'+2Q'=\Phi'=0$ and $P'=Q'=-A'$, the components of the Ricci tensor can also be calculated in terms of $A$ and $T$ directly using the metric given by (\ref{metric-1a}), and the components are
\ba
\frac{R_{\mu\nu}}{G_{\mu\nu}}&=&-\frac{R_{mn'}}{G_{mn'}}=
-\frac{1}{r_s^2}\,e^{-2T}
(A''-A'^2-T'A'),\nn\\ R_{tt}&=&\frac{1}{r_s^2}\,e^{-2T}
(A''-9A'^2-T'A')\,G_{tt}.\label{RMN-3}
\ea
Using (\ref{RMN-2}) and (\ref{RMN-3}), we determine $T$ in terms of $u$,
\be
e^{2T}=(1+2\sech 2u)\sech u.\label{Tsoln-1}
\ee

At this point all metric and flux components are expressed in terms of a single variable $u$ and it is straightforward to verify that the solutions indeed solve all the supergravity equations.
Notice that all components of the solutions, except $T$ which sets the definition of the radial coordinate, were obtained from the supersymmetry equations of motion with the Bianchi identities imposed on them.

For completeness, the right hand side of (\ref{phi-1}) vanishes and the dilaton is constant.
The left hand side of (\ref{RMN-1}) obtained using the metric and given by (\ref{RMN-3}) and the right hand side obtained using the fluxes and given by (\ref{RMN-2}) are identical for each and every component.
We have already imposed the Bianchi identities $dF_3=-F_1\wedge H_3$ and $d\tilde{F}_5=H_3\wedge F_3$.
The remaining equations are
\ba
&d\star_{10}\left(e^{2\Phi}F_1\right)=-e^{2\Phi}H_3\wedge \star_{10}F_3,\quad d\star_{10}\left(e^{2\Phi}F_3\right)=F_5\wedge H_3,&\nn\\ &d\star_{10}\left(H_3-e^{2\Phi}C_0 F_3\right)=-e^{2\Phi}F_5\wedge F_3.&\label{bosonic-2}
\ea
Both sides of the first two equations vanish and the solutions also satisfy the last equation in (\ref{bosonic-2}) with both the left and the right hand sides being equal to $r_s^2\,(\frac{g_sN}{2\pi})^2$ $\sqrt{1+2\sech 2u}$ $\cosh 2u \sinh u \,(dx^0\wedge dx^1 \wedge dx^2 \wedge dx^3 \wedge dt \wedge d\psi \wedge d\varphi_1 \wedge d\varphi_2 +dx^0\wedge dx^1 \wedge dx^2 \wedge dx^3 \wedge dt \wedge d\psi \wedge d\varphi_3 \wedge d\varphi_4)$.

Thus the solutions solve each and every one of all supergravity equations.\footnote{This also provides a nontrivial check on the correctness of the equations of motion we wrote in \cite{Hailu:2007ae} with all flux and torsion components in the subtle $3\oplus\bar{3}$ sector turned on.}

\section{Radial coordinate\label{sec-radial}}

Now we establish the relation between the variable $u$ in terms of which the supergravity solutions are expressed and the radial coordinate $r$ that describes actual distances on $\mathds{R}^1$. First we rewrite the metric (\ref{metric-1a}) using (\ref{PQA-soln-2}) and (\ref{Tsoln-1}),
\ba
ds_{10}^2&=&\cosh u\,dx_{1,3}^2+r_s^2\,\sech u\,\Bigl((1+2\sech 2u)\,dt^2\nn\\&&+d\psi^2+d\varphi_1^2+d\varphi_2^2
+d\varphi_3^2+d\varphi_4^2 \Bigr).\label{metric-2aa}
\ea
The radial coordinate $r$ is defined by
\be
\frac{3\,dr}{r}=\sqrt{1+2\sech 2u}\,dt.\label{rut-1}
\ee
We know the relation between $u$ and $t$ from (\ref{u-soln-1}) which with (\ref{ephi-1}), (\ref{PQA-soln-2}), and (\ref{Tsoln-1}) gives
\be
\tanh u= \frac{3g_sN}{2\pi}\,\ln(\frac{r}{r_s})= \frac{g_sN}{2\pi}\,\rho.\label{ur-1}
\ee
Notice that the range of $r$ is finite with
\be
r_s\le r\le r_s\, e^{\frac{2\pi}{3g_sN}}\label{r-range-1}
\ee
and the background geometry is compact.

\section{Summary of solutions\label{sec-solsum}}

In this section the complete set of supergravity solutions obtained in sections \ref{sec-metric-flux}-\ref{sec-radial} for the background of the 4D gauge theory are summarized in one place for convenience and some of their features are analyzed. The metric is
\ba
ds_{10}^2&=&\cosh u\,dx_{1,3}^2+r_s^2\,\sech u\,\Bigl((1+2\sech 2u)\,dt^2\nn\\&&+d\psi^2+d\varphi_1^2+d\varphi_2^2
+d\varphi_3^2+d\varphi_4^2 \Bigr).\label{metric-2}
\ea
The dilaton is constant,
\be e^\Phi=g_s.\label{ephi-1aa}\ee
The fluxes are
\ba
&F_1=\frac{N}{2\pi}d\psi,&\nn\\ &F_3=\frac{N}{2\pi}\,r_s^2\,\tanh u \, \left(d\psi\wedge d\varphi_1\wedge  d\varphi_2 +d\psi\wedge  d\varphi_3\wedge  d\varphi_4\right),&\nn\\ &H_3=\frac{g_sN}{2\pi}\,r_s^2\,\sqrt{1+2\cosh 2u}\left(dt\wedge d\varphi_1\wedge  d\varphi_2+dt\wedge d\varphi_3\wedge  d\varphi_4\right), &\nn\\
&\tilde{F}_5=\frac{N}{2\pi}\,r_s^4\,\left(1+\tanh^2u\right) \, d\psi\wedge d\varphi_1\wedge  d\varphi_2\wedge  d\varphi_3\wedge  d\varphi_4&\nn\\ &-\frac{N}{2\pi}\,\sqrt{1+2\cosh 2u}\,\cosh 2u\, \cosh^2u \, d^4x \wedge dt.
\qquad \,& \label{F5sol-1bb}
\ea
Notice that the only R-R flux that is closed is $F_1$ as it is the only one that is directly sourced  by the D7-branes, the others are induced by the backreaction of the background and satisfy the Bianchi identities.

The variable $u$ is related to the radial coordinate $r$ by $\tanh u= \frac{3g_sN}{2\pi}\ln(\frac{r}{r_s})= \frac{g_sN}{2\pi}\,\rho$, $0\le \rho\le \frac{2\pi}{g_sN}$ and $0\le u\le \infty$.
The volume element in 10D is
$r_s^{6}\,\sech u$ $\sqrt{1+2\sech 2u}\,dx^0\wedge dx^1\wedge dx^2\wedge
dx^3\wedge dt\wedge d\psi \wedge d\varphi_1\wedge d\varphi_2\wedge d\varphi_3\wedge d\varphi_4$, which is finite and nonzero at $\rho=0$ and vanishes at $\rho=\frac{2\pi}{g_sN}$.
The background is compact
and the volume coming from the extra space is finite.
In terms of the radial coordinate $r$, $r_s\le r\le r_{b}$, where $r_{b}=r_s\,e^{\frac{2\pi}{3g_sN}}$. The range of $r$ is smaller for larger $g_sN$.
We call the boundary at $r=r_s$ the IR boundary and that at $r=r_b$ the UV boundary in anticipation of their relation to the energy scale in the gauge theory.
At the UV boundary,  $u=\infty$, the warp factor on $\mathds{R}^{1,3}$ is infinite and the radius of $\mathds{S}^1\times \frac{\mathds{T}^2}{Z_2}\times \frac{\mathds{T}^2}{Z_2}$ vanishes and we have 4D spacetime. Notice that the compactness of the extra dimensional space comes with a singularity at the UV boundary. The region of interest to us is the IR region, and supergravity flow is smooth in the IR.
The radius of $\mathds{S}^1\times \frac{\mathds{T}^2}{Z_2}\times \frac{\mathds{T}^2}{Z_2}$ is largest in the IR and equals $r_s$ at the IR boundary.

The complex coordinates on $\mathds{C}^1$ and the two $\mathds{T}^2$'s discussed in section \ref{sec-metricansatz} can be written as $z^1=(\frac{r}{r_s})^3\,e^{i\psi}=e^{\rho+i\psi}$, $z^2=\varphi_1+i\varphi_2$, and $z^3=\varphi_3+i\varphi_4$.
All the $F_1$, $F_3$, $\tilde{F}_5$, and $H_3$ fluxes transform with charges $(+,\,+,\,+)$ for $(g_1,\,g_2,\,g_3)$ under $Z_2\times Z_2\times Z_2$ and the supergravity relations obey the discrete orbifold symmetry.

The metric on $\mathds{R}^{1,3}\times \frac{\mathds{C}^1}{Z_2}\times \frac{\mathds{T}^2}{Z_2}\times \frac{\mathds{T}^2}{Z_2}$ given by (\ref{metric-2}) is rewritten in terms of the radial coordinate $r$ as
\be
ds_{10}^2={\cosh u}\,dx_{1,3}^2
+{r_s^2}\, \sech u\,\left(9\,\frac{dr^2}{r^2}+d\psi^2+d\varphi_1^2+d\varphi_2^2
+d\varphi_3^2+d\varphi_4^2\right),\label{metric-3}
\ee
where
\be
\sech u= \sqrt{1-(\frac{g_sN}{2\pi}\rho)^2}=\sqrt{1-(\frac{3g_sN}{2\pi}
\ln(\frac{r}{r_s}))^2}\,.
\label{h1}
\ee
The compactified 6D space is complex, non-K\"ahler, and conformally Calabi-Yau.\footnote{We pointed out in \cite{Hailu:2007ae} that flows on conformally Calabi-Yau backgrounds with nonzero components of $F_3$ and $H_3$ fluxes in the $6\oplus\bar{6}$ representation have constant axion-dilaton coupling. Here, the 3-form fluxes are all nonprimitive, all nonzero flux and torsion components are in the $3\oplus\bar{3}$ representation, and the axion-dilaton coupling is not constant.} One way to see that is to note that the torsion components have the relation $3W_4=2W_5$ with the other components vanishing.

The metric is smooth at the IR boundary $\rho=0$ (or $r=r_s$). The singularity at $\rho=0$ in the background of the unwrapped D7-branes is removed by the fluxes that are induced when the D7-branes are wrapped over the 4-cycle.

The worldvolume element of D7-branes at $r$ is $dV_{D7}=r_s^4\,d^4x\,d\varphi_1\,d\varphi_2\,d\varphi_3\,d\varphi_4$ which is  independent of $r$, and the wrapped D7-branes stay put at $r=r_s$. Where on $\mathds{S}^1$ the D7-branes are located is related to chiral symmetry breaking which will be discussed in section \ref{sec-csb}. The radius of $\mathds{S}^1\times \frac{\mathds{T}^2}{Z_2}\times \frac{\mathds{T}^2}{Z_2}$ at $\rho$ is $R_5=r_s\sqrt{\sech{u}}=r_s(1-(\frac{g_sN}{2\pi}\rho)^2)^{\frac{1}{4}}$.

\begin{figure}[t]
\centering
\subfloat[Radius of $\mathds{S}^1\times \frac{\mathds{T}^2}{Z_2}\times \frac{\mathds{T}^2}{Z_2}$.]{\label{fig:rad5}
\includegraphics[width=0.5\textwidth]{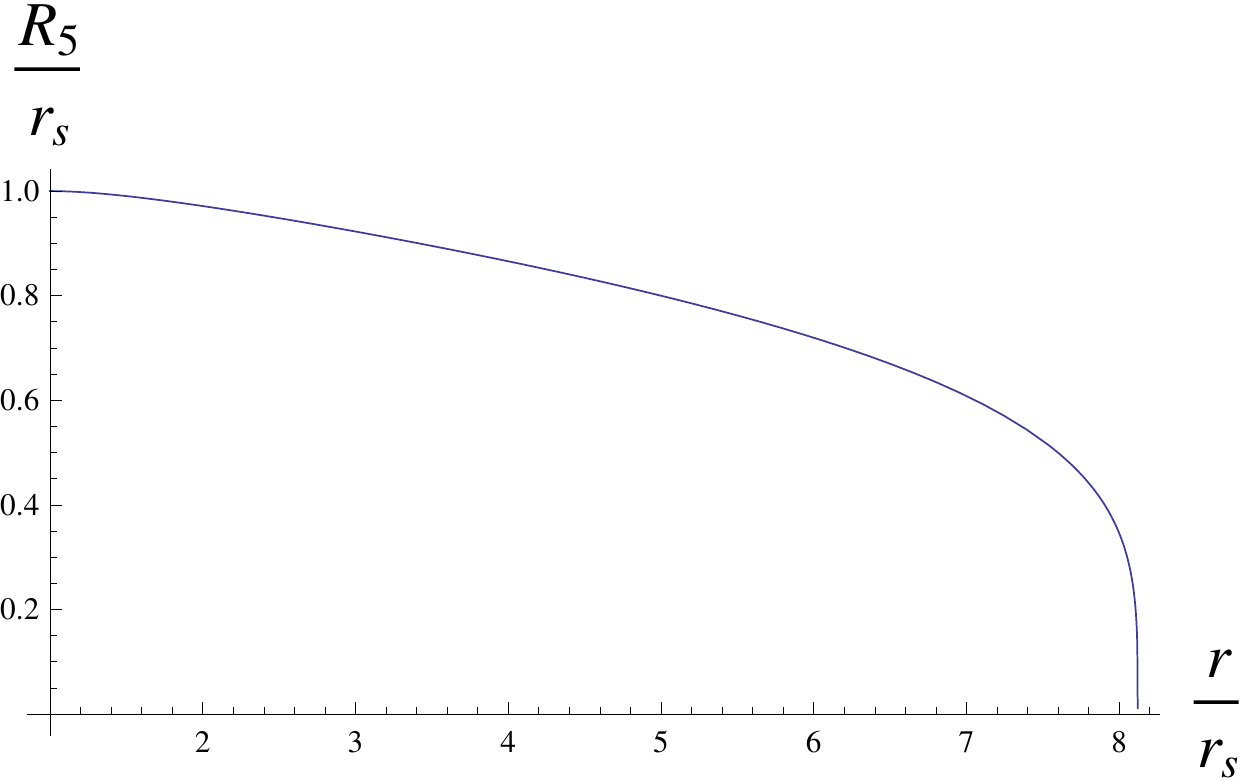}} \hspace{5mm}
\subfloat[Warp factor.]{\label{fig:w4dr}\includegraphics[width=0.515\textwidth]{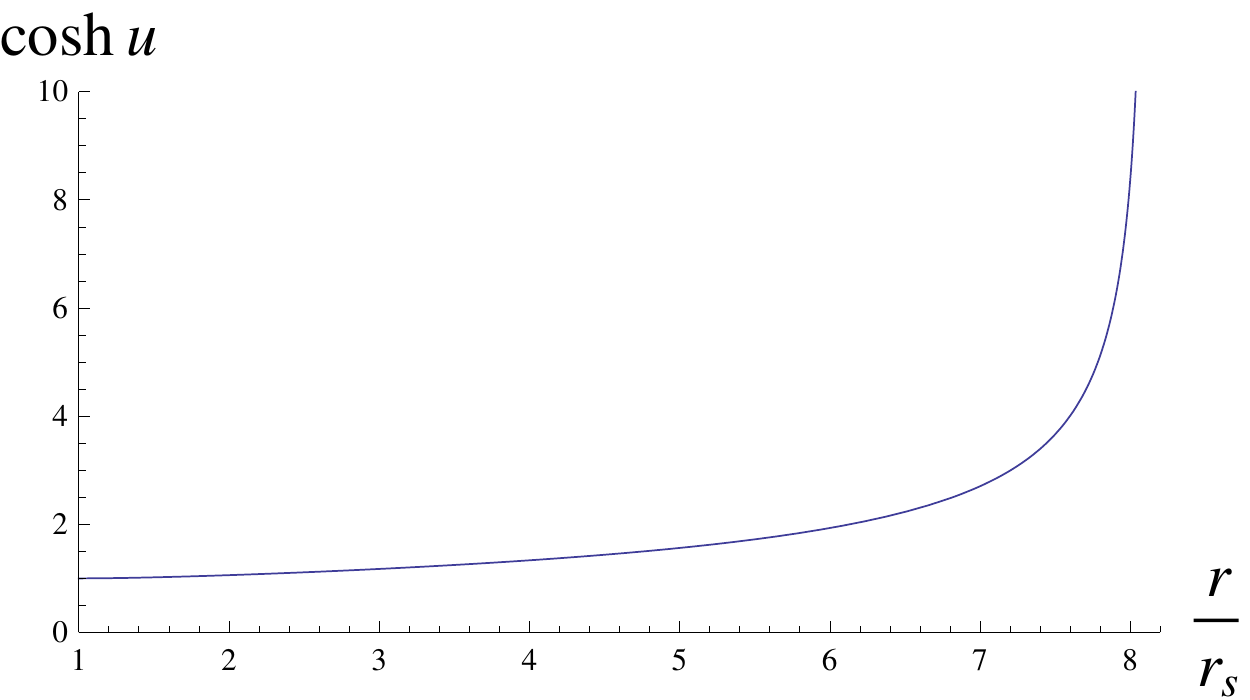}}
\caption{Radius and warp factor  for $g_sN=1$. The range of $r$ and the region where the curvature is nearly constant can be made larger by taking smaller $g_sN$.}\label{fig:backg}
\end{figure}

Figure \ref{fig:rad5} shows the radius of $\mathds{S}^1\times \frac{\mathds{T}^2}{Z_2}\times \frac{\mathds{T}^2}{Z_2}$ and Figure \ref{fig:w4dr} shows the warp factor on the 4D  spacetime, both plotted using (\ref{h1}) for the 't Hooft coupling at the UV boundary $g_sN=1$. The UV boundary is at $r=e^{\frac{2\pi}{3}}\,r_s\approx 8\,r_s$. The range of $r$ and the region where the curvature is nearly constant can be made larger by taking smaller $g_sN$. For instance, the UV boundary is located at $r\approx 66\, r_s$ for $g_sN=0.5$ and at $r\approx1.2\times10^9\, r_s$ for $g_sN=0.1$.

\section{Background geometry in diagrams\label{sec-figs}}

\begin{figure}[t]
\centering
\subfloat[Background geometry of 8D gauge theory.]{\label{fig:backg8Da}\includegraphics[width=0.47\textwidth]
{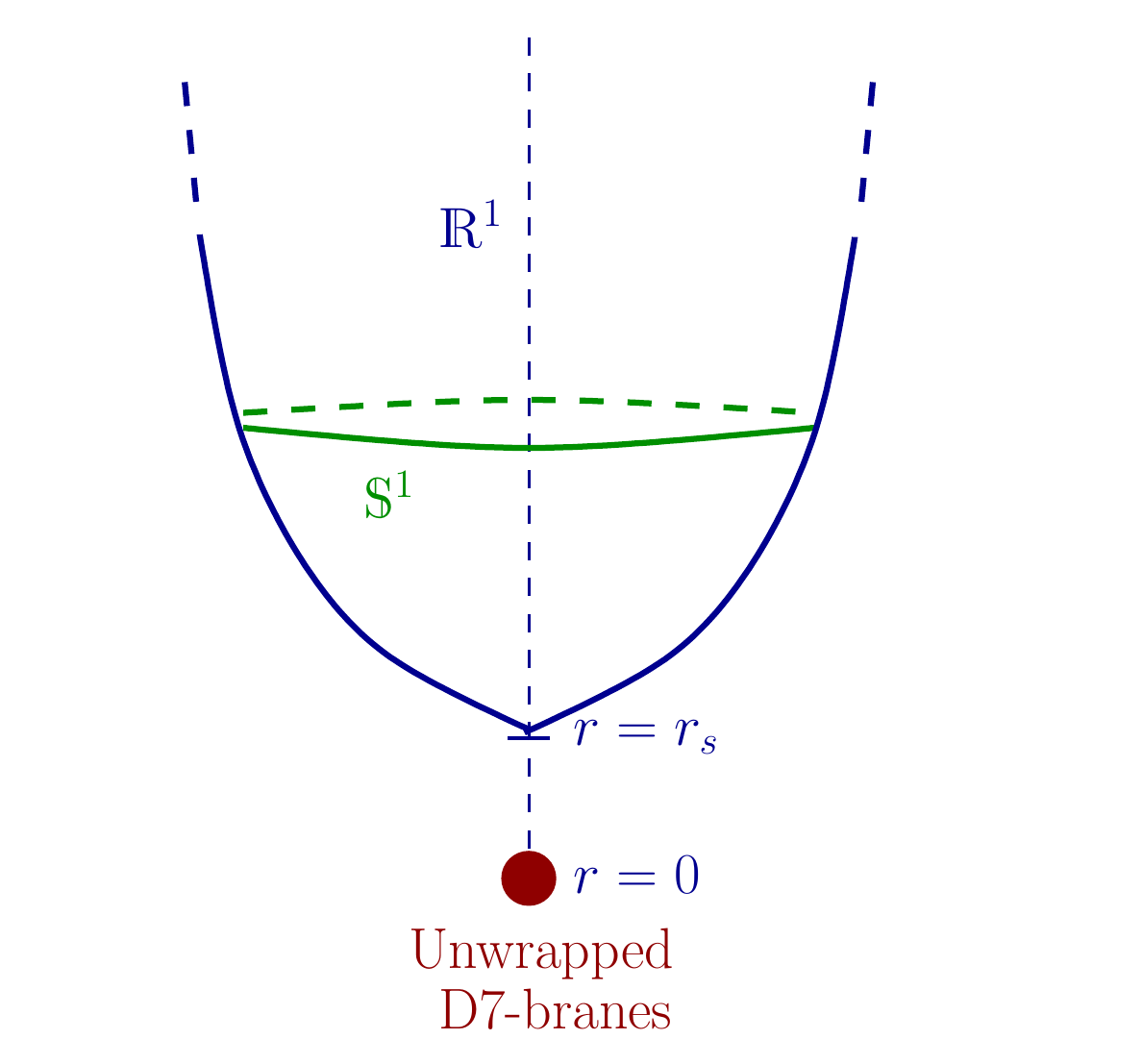}}
\hspace{3mm}
\subfloat[Background geometry of 4D gauge theory.]{\label{fig:backg4Da}\includegraphics[width=3.10in]{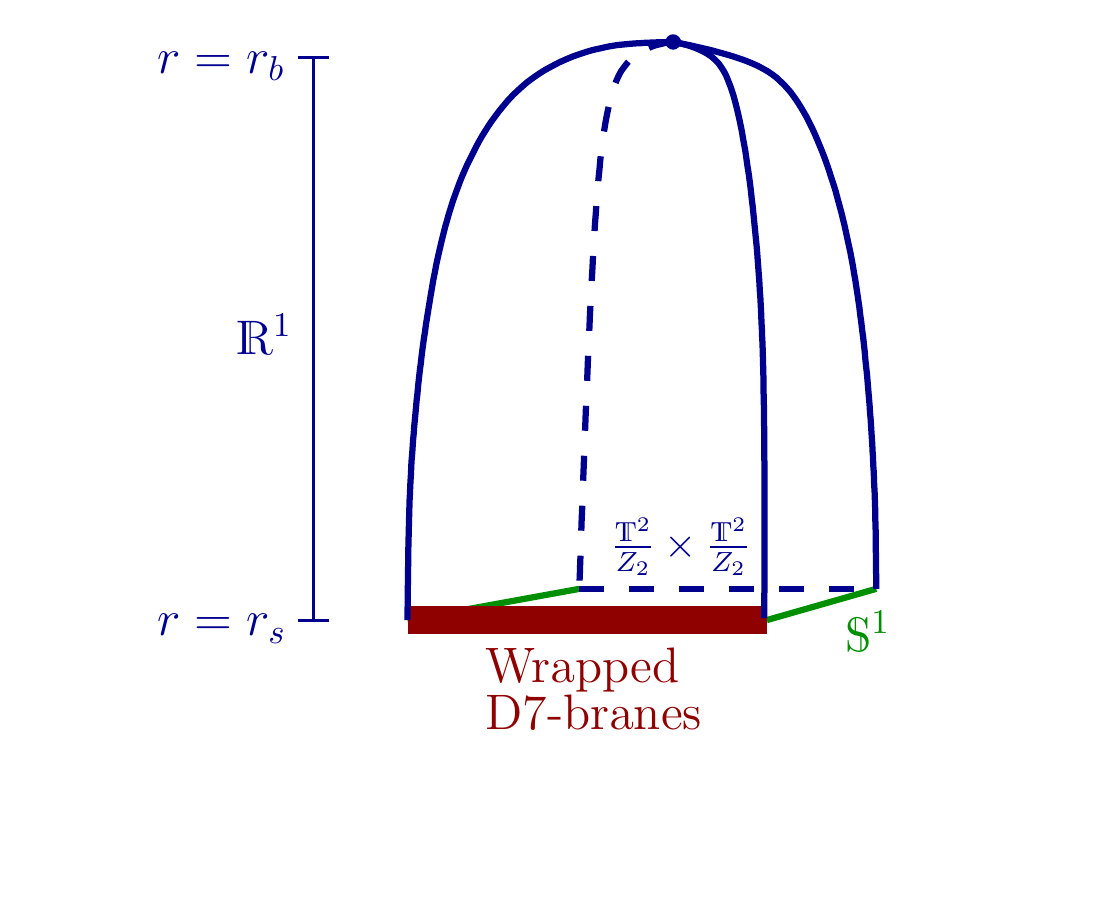}} \caption{Schematic representation of the background geometry.}\label{figs:schem}
\end{figure}

Diagrams that show a schematic representation of the background geometry are given in Figure \ref{figs:schem} with $\mathds{R}^1$ parameterized by $r$.
Figure \ref{fig:backg8Da} schematically shows the background geometry of the 8D gauge theory with unwrapped D7-branes which fill flat 8D spacetime $\mathds{R}^{1,7}$ at the singularity point of the $\frac{\mathds{C}^1}{Z_2}$ cone at $r=0$. The $F_1$ flux and the dilaton are turned on with the $F_1$ flux wrapping around $\mathds{S}^1$ and the string coupling $e^\Phi$ diverging at $r=r_s$. The metric on $\mathds{R}^1\times\mathds{S^1}$ is warped by a factor of $e^{2H}=\rho$, which vanishes with $\mathds{S^1}$ having zero-size and spacetime being 8D at $r=r_s$ or $\rho=0$, and the physical region of  the supergravity flow is cutoff to $r\ge r_s$. The range of $\psi$ on $\mathds{S^1}$ in our normalization is $[0,2\pi]$; an independent region of $[0,\pi]$ can be chosen with the boundary condition $\psi\sim \psi+\pi$ imposed.

The D7-branes at the singularity are then wrapped over $\frac{\mathds{T}^2}{Z_2}\times \frac{\mathds{T}^2}{Z_2}$ of radius $r=r_s$ while filling flat 4D spacetime $\mathds{R}^{1,3}$. Now all $F_1$, $F_3$, $H_3$, and $F_5$ fluxes are turned on and the metric on $\mathds{S}^1\times\frac{\mathds{T}^2}{Z_2}\times \frac{\mathds{T}^2}{Z_2}$ is warped by a factor of $\sech u= \sqrt{1-(\frac{3g_sN}{2\pi}\,\log{(\frac{r}{r_s}}))^2}$, which is finite at $r=r_s$ and $\mathds{S}^1$ is blown-up, and vanishes at $r=r_b=r_s\,e^{\frac{2\pi}{3g_s N}}$. Spacetime is $\mathds{R}^{1,3}\times\mathds{S}^1\times\frac{\mathds{T}^2}{Z_2}\times \frac{\mathds{T}^2}{Z_2}$ and the background geometry has largest size at the IR boundary at $r=r_s$. At the UV boundary at $r=r_b$, $\mathds{S}^1\times\frac{\mathds{T}^2}{Z_2}\times \frac{\mathds{T}^2}{Z_2}$ is warped to zero-size and spacetime is 4D. Figure \ref{fig:backg4Da} schematically shows the background geometry of the 4D gauge theory with wrapped D7-branes.

Notice that what we have here is different from standard geometric transitions such as in \cite{Klebanov:2000hb} and \cite{Maldacena:2000yy} where cycles wrapped by branes are blown-down, the branes disappear, and other cycles are blown-up by fluxes. Here the 4-cycle on $\frac{\mathds{T}^2}{Z_2}\times \frac{\mathds{T}^2}{Z_2}$ that is wrapped by the D7-branes at the singularity point on the background of the unwrapped D7-branes has not disappeared and $\mathds{S}^1$ is blown-up by flux, and both the 4-cycle and $\mathds{S}^1$ are warped to zero-size at the UV boundary. The singularity at $\rho=0$ before the D7-branes are wrapped and the singularities at the fixed points of the $\frac{\mathds{T}^2}{Z_2}\times \frac{\mathds{T}^2}{Z_2}$ orbifold are smoothed out by flux, since the fluxes have nonzero components through them. There is one singularity located at the UV boundary in the final geometry where $\mathds{S}^1\times\frac{\mathds{T}^2}{Z_2}\times \frac{\mathds{T}^2}{Z_2}$ is warped to zero-size.

\section{Four-dimensional gauge coupling\label{sec-4dgc}}

The action for the 4D gauge theory that lives on the D7-branes wrapped over the 4-cycle at $t=0$ is, using the metric given by (\ref{metric-2}),
\be
S_{4}=\frac{1}{g_8^2} V_4 \int d^4x\, \sqrt{-g}\,g^{\mu\sigma}g^{\nu\delta}F_{\mu\nu}F_{\sigma\delta},
\label{sd7-1}
\ee
where $g_{\mu\nu}$ is the metric, $F_{\mu\nu}$ is the Yang-Mills field strength, and $x^\mu$ are the coordinates on $\mathds{R}^{1,3}$, $V_4$ is the volume of the 4-cycle, and $g_8$ is given by (\ref{g8-1}). The gauge coupling of the 4D gauge theory on the wrapped D7-branes is related to the gauge coupling of the 8D gauge theory on the unwrapped D7-branes and the volume of the wrapped 4-cycle at $t=0$ and, using (\ref{g8-1}) and (\ref{metric-3}),
\be
\frac{4\pi}{g_4^2}=\frac{4\pi}{g_8^2} V_{T_4}=\frac{3N}{2\pi}\,\log(\frac{r}{r_s})\,\frac{r_s^4}{\alpha'^2}.
\label{g4-1-2a}
\ee

Now we identify
\be r_s^2=\alpha'\,;\label{rsal-1}\ee
the only independent parameter of the string theory, $\alpha'$, is identified with the only dimensionful parameter in the gravity background we have constructed here, $r_s$. With this, (\ref{g4-1-2a}) reduces to
\be
\frac{4\pi}{g_4^2}
=\frac{3N}{2\pi}\,\log(\frac{r}{r_s})=\frac{N\rho}{2\pi}.
\label{g4-1-2}
\ee
Notice that $\frac{4\pi}{g_4^2}=e^\Phi$, the string coupling in the background of the 8D gauge theory. The background before the branes are wrapped has induced a running dilaton in addition to the $F_1$ flux and the background with wrapped branes has constant dilaton with all fluxes of type IIB theory induced. The gauge coupling of the 4D gauge theory on the wrapped D7-branes inherits the running dilaton from the background before the branes are wrapped.

The 3-form fluxes summarized in section \ref{sec-solsum} do not form imaginary self-dual or imaginary anti-self-dual combination.
Now we notice that the 3-from fluxes in the background of the 4D gauge theory with the running of the dilaton inherited from the background of unwrapped branes actually form an imaginary anti-self-dual combination. Consider the combination of 3-form fluxes $G_3=F_3-\frac{i}{g_s}\,\tanh u\, H_3$. Taking the Hodge star, we have $\star_6 G_3=-i\,G_3$. Therefore, $G_3$ is imaginary anti-self-dual, where $G_3$ has the standard form of anti-self-dual 3-form flux $F_3-i\,e^{-\Phi} \,H_3$ with $g_s\,e^{-\Phi}=\tanh u=\frac{g_sN\rho}{2\pi}$ in our notation, where $\Phi$ is the dilaton in the background of the 8D gauge theory on the unwrapped D7-branes given by (\ref{phi-T-8d-1}).

Next we need to find the relation between the scale of the gauge theory, $\Lambda$, and the radial coordinate in the gravity theory, $r$. We do that, following \cite{Maldacena:1998re}, by considering the energy of a string that is stretched between the D7-branes at the IR boundary at $r=r_s$ and probe branes at some $r$ between the IR and the UV boundaries.
We are interested in the energy of the string in static configuration at time $X^0\to \infty$ and we parameterize the string worldsheet in terms of $\sigma^0=x^0$ and $\sigma^1=r$.
The worldvolume action of the string, using the metric given by (\ref{metric-3}), with all spacetime intervals (including those involving 4D spacetime and $r$) in this paragraph measured in units of $r_s=\sqrt{\alpha'}$ is
\be
S_2=\frac{1}{2\pi\alpha'}\int d\sigma^0 d\sigma^1\, \sqrt{\det G_{MN}\,\partial_{a}X^M \partial_{b}X^N}=\frac{X^0}{2\pi}\,\ln\, {r}^3,
\label{S2-1}
\ee
where $a$ and $b$ are indices for the coordinates on the two-dimensional string worldsheet.

Thus the $r$ dependence of the energy of the string is, putting $r_s$ back,
\be E\sim \ln\, \left(\frac{r}{r_s}\right)^3 \label{Er-1}\ee
which is similar to the $\Lambda$ dependence of the Veneziano-Yankielowicz superpotential if we map the scale of the gauge theory and the radial coordinate in the gravity theory as
\be
\frac{\Lambda}{\Lambda_s} \sim \frac{r}{r_s}.\label{ggmapping-1}
\ee

Notice that $\Lambda_s\sim r_s$; the only dimensionful parameter in the gauge theory, the nonperturbative scale $\Lambda_s$, is mapped to the only dimensionful parameter in the gravity theory, the radius $r_s$ of $\mathds{S}^1\times \frac{\mathds{T}^2}{Z_2}\times \frac{\mathds{T}^2}{Z_2}$ at the IR boundary.
With (\ref{ggmapping-1}), (\ref{g4-1-2})
becomes $\frac{4\pi}{g_4^2}= \frac{3N}{2\pi}\,\log(\frac{\Lambda}{\Lambda_s})$, which
reproduces the renormalization group flow of the 4D gauge coupling given by (\ref{dTdmu-1}).

We can also see the appearance of the dimensionful parameter in 4D (or the dimensional transmutation of the gauge coupling) in the gravity theory by rescaling the metric given by (\ref{metric-3}) such that $\sech u\to \frac{\sech u}{r_s^2}$ which gives
\be
ds_{10}^2={r_s^2}\,{\cosh u}\,dx_{1,3}^2
+ \sech u\,\left(9\,\frac{dr^2}{r^2}+d\psi^2+d\varphi_1^2+d\varphi_2^2
+d\varphi_3^2+d\varphi_4^2\right).\label{metric-3bb}
\ee
Thus ${r_s}$ sets the mass scale for an observer in 4D and $\Lambda_s\sim r_s$.

\section{Analysis of couplings and curvature\label{sec-coup-curv}}

In this section we analyze the magnitudes of the string, the gauge, and the 't Hooft couplings and the curvature of the compact space and discuss the validity of the supergravity description of the string theory.

The 't Hooft coupling, $\frac{g_4^2N}{4\pi}=e^{\Phi}N$, is the parameter which determines whether the gauge theory is perturbative or nonperturbative. For $e^{\Phi}N<1$, perturbative gauge theory is applicable. When $e^{\Phi}N>>1$, the gauge theory is highly nonperturbative, and our interest is in a gravity description of the 4D gauge theory in this region. The magnitude of the string coupling, $e^\Phi$, determines whether quantum loop corrections to the classical supergravity are necessary or negligible. When the curvature of the background is large ($>\frac{1}{\sqrt{\alpha'}})$, $\alpha'$ corrections to the supergravity description are relevant. When $N>>1$, only 't Hooft's ribbon Feynman graphs on a sphere (planar diagrams) are relevant. When both $e^{\Phi}N$ and $N$ are large, we have planar ribbon graphs which become dense at higher order in $e^{\Phi}N$ and which are expected to correspond to string worldsheets in the string theory.

In our case, first the constant string coupling in the background of the 4D gauge theory, which corresponds to the string coupling at $\rho= \frac{2\pi}{g_s N}$ in the background of the 8D gauge theory, can be taken to be small, $g_s<<1$, such that string loop corrections are small.
The magnitude of the string coupling in the background of the 8D gauge theory which is also inherited by the 4D gauge theory is $e^\Phi=\frac{2\pi}{N\rho}$ and $g_s\le e^{\Phi}\le 1$ for $\frac{2\pi}{N}\le \rho \le \frac{2\pi}{g_s N}$ which covers the whole range of $\rho$ for $N\to \infty$. Because the gauge coupling diverges at $\rho=0$ the supergravity flow requires a regularization with an IR cutoff at some $\rho=\rho_0>0$, see \cite{Hailu:2011conf} for further discussion. At $\rho\ge\rho_0\ne 0$, $e^{\Phi}$ has maximum value of $\frac{2\pi}{N\rho_0}$, and we can make $e^{\Phi}<<1$ by making $N$ appropriately large. Thus string loop corrections are small for the whole range of $\rho$ for large $N$. The 't Hooft coupling is $\frac{2\pi}{\rho}\ge g_s N$ for the range of $\rho$ we have here and its value is smallest and equals $g_s N$ at the UV boundary. Notice that the 't Hooft coupling is always large ($>>1$) in the region of interest in the IR.

Next let us analyze the magnitude of the curvature, which we can do by looking at the components of the Ricci tensor given by (\ref{RMN-2}). The Ricci scalar is $R=\frac{1}{r_s^2}(\frac{g_s N}{2\pi})^2 \cosh^3 u=\frac{1}{r_s^2}(\frac{g_s N}{2\pi})^2 \frac{1}{(1-(\frac{3g_sN}{2\pi}\,\log{(\frac{r}{r_s}}))^2)
^{\frac{3}{2}}}$. The curvature depends on the location on the radial direction. All components of the Ricci tensor are small, $<<\frac{1}{\alpha'}$, in the IR region for $\frac{g_sN}{2\pi}<<1$. Recall that $g_sN$ is the 't Hooft coupling at the UV boundary which can be made very small. Thus the curvature is small in the IR region where the gauge theory is strongly is coupled and a dual gravity description is useful. The reason we have managed to get small curvature is because the range of the radial direction is inversely related to the magnitude of the 't Hooft coupling at the UV boundary. The curvature becomes large near the UV boundary and diverges at the boundary where the value of $g_sN$ can be chosen such that the appropriate description is the gauge theory.
The range of the radial direction is $r_s\le r\le r_s e^{\frac{2\pi}{3g_s N}}$ and its size is $\Delta r=(e^{\frac{2\pi}{3g_s N}}-1)r_s>rs=\sqrt{\alpha'}$ for $g_sN\le \frac{2\pi}{3\ln 2}\approx3$. For instance, for $g_s N\approx 1$, $\Delta r \approx 7\, r_s$. The magnitude of $\Delta r$ can be made large by taking $g_sN$ small and becomes infinite for $g_sN=0$; it is smaller for larger $g_sN$, since the strength of gauge coupling running increases and the theory flows faster to a region where the gauge theory is weakly coupled.  A choice of $g_sN\approx 1$ or smaller provides a UV cutoff beyond which perturbative gauge theory description is appropriate. In the IR region away from the UV boundary, where the 't Hooft coupling is large and the curvature is smallest, the classical gravity description is appropriate.
This is different from previous constructions of gauge/gravity duality, including \cite{Maldacena:1998re, Klebanov:2000hb, Maldacena:2000yy}, where the curvature is smaller for larger 't Hooft coupling and the space transverse to the radial direction vanishes as $g_sN\to 0$. The curvature is constant in \cite{Maldacena:1998re} and it is largest in the IR and gets smaller in the UV in \cite{Klebanov:2000hb} and \cite{Maldacena:2000yy}.

Now let us consider a quark and an antiquark located at the UV boundary or at some other $r$ in 4D with a string between them. As it can be seen from the metric, and it is discussed further in \cite{Hailu:2011conf}, the string is stretched toward the IR boundary and the tension of the string at the location of the wrapped D7-branes at $r=r_s$ is $T_s\sim \frac{1}{2\pi r_s^2} =\frac{1}{2\pi\alpha'}$. This is the string tension that is seen by a 4D observer. KK modes and glueballs on $\mathds{S}^1\times \frac{\mathds{T}^2}{Z_2}\times \frac{\mathds{T}^2}{Z_2}$ at the IR boundary where $r=r_s$ come with mass of magnitude $m_{\mathrm{KK}}^2 \gtrsim \frac{1}{r_s^2}=\frac{1}{\alpha'}$.
Thus the string tension is of the same order and  smaller than the square of KK or glueball masses.
Because the compactified space contains only product of $\mathds{S}^1$'s, a truncation of the KK spectrum to the zero-modes is consistent.
For contrast, the string tension is much bigger than the KK masses in previous gauge/gravity constructions, such as in \cite{Klebanov:2000hb} and \cite{Maldacena:2000yy}, with larger separation for larger 't Hooft coupling.
That is because the tension measured by a 4D observer increases with increasing magnitude of warp factor while the KK masses decrease with increasing size of internal space.

The Ricci scalar is positive and the 10D spacetime is positively curved. The components of the Ricci tensor in the 4D space are $R_{\mu\nu}=\frac{1}{4\,r_s^2}(\frac{g_sN}{2\pi})^2 \cosh 2u\,
\cosh^2 u (1-(1+2\sech 2u)$ $\cosh 2u)\eta_{\mu\nu}$, where $\eta_{\mu\nu}$ is flat Minkowski metric. The magnitude of the components is negative for the whole range and $R_{\mu\nu}=-\frac{1}{2\,r_s^2}\left(\frac{g_sN}{2\pi}\right)^2
\eta_{\mu\nu}$ at the IR boundary. Thus the warp factor results in a negatively curved 4D spacetime. The components in the extra dimensional space are positive with $R_{mn}=\frac{1}{2\,r_s^2}\left(\frac{g_sN}{2\pi}\right)^2
G_{mn}$ at the IR boundary.
The volume of the 10D spacetime follows from the metric given by (\ref{metric-2bb}), $V_{10}=V_4\, r_s^6 (2\pi)^5$ $\int_{0}^{\frac{2\pi}{g_sN}} \sqrt{1-(\frac{g_sN}{2\pi}\,\rho)^2}\,d\rho=V_4\,V_6$, where $V_4$ is the volume of the 4D spacetime and $V_6 =  \frac{16\pi^7}{g_sN} r_s^6= \frac{16\pi^7}{g_sN} \alpha'^3$ is the volume coming from the extra dimensional compact space. Thus $V_6>>\alpha'^3$ for $g_sN << 16\pi^7$ and it is finite. A small and finite Newton's constant in 4D can be obtained from a large one in 10D with appropriately small $g_sN$ which gives large $V_6$.

The curvature of the background of the 8D gauge theory on the unwrapped D7-branes is large in the IR region near $r=r_s$. However, the only quantity in the background of the 4D gauge theory that is inherited from the 8D gauge theory background is the gauge coupling which agrees with the exact nonperturbative gauge coupling running of the gauge theory. Therefore, we do not expect $\alpha'$ corrections to modify it.

Spacetime is effectively 4D for an observer in $\mathds{R}^{1,3}$ who measures interactions between particles (say quarks) separated by a distance $>>r_s$ on $\mathds{R}^{1,3}$.

In conclusion, the background is such that the classical supergravity limit of the string theory with appropriately large $N$ and small $g_sN$ accommodates physically interesting and radially varying 't Hooft coupling that is large, the gauge theory is strongly coupled, and the curvature is small in the IR region where a dual gravity description is useful.

\section{Chiral symmetry breaking\label{sec-csb}}

The $U(1)$ R-symmetry corresponds to rotations on $\mathds{S}^1$ in the internal space normal to the D7-branes, which shift $\psi$ by a real number. With the axion potential given by (\ref{F1C0-1}), we see that the coordinate $\psi$ in the gravity theory is mapped to the Yang-Mills angle $\Theta$ in the gauge theory as \be
N\psi=\Theta.\label{Theta-1}
\ee
Notice that the 4D gauge coupling and the Yang-Mills angle combine into the coupling coefficient
\be \tau=\frac{4\pi i}{g_4^2}+\frac{\Theta}{2\pi}=\frac{N}{2\pi}\left(i\,\rho
+\psi\right),
\label{tau-1}\ee
which is anti-holomorphic function, with $\bar{z}^{\bar{1}}=\rho-i\psi$ in our notation.
Worldsheet instantons wrapping $\frac{\mathds{T}^2}{Z_2}\times \frac{\mathds{T}^2}{Z_2}$ give phase $\Theta$ and the $U(1)$ symmetry is anomalous.
Because $N\psi\sim N\psi+\pi$ in the $\frac{\mathds{C}^1}{Z_2}$ orbifold, there is a reduced anomaly-free discrete $Z_{2N}$ symmetry,  $\psi\to \psi+\frac{2\pi k}{2N}$, $k=1,2,\cdots,2N$, that survives. This is the same as the anomaly-free $Z_{2N}$ R-symmetry of quantum gauge theory.

The $Z_2$ symmetry in the $\frac{\mathds{C}^1}{Z_2}$ orbifold involves ${\psi} \sim {\psi} +\pi$ which reduces the $Z_{2N}$ symmetry down to $Z_2$ giving rise to $N$ discrete vacua.
The $N$ discrete vacua are represented by $\psi_m=\frac{2\pi m}{N}$, with  $m=0,\cdots,N-1$, or the corresponding discrete values for the axion potential $(C_0)_m=m$.
Thus the gravity theory reproduces the pattern of chiral symmetry breaking of the gauge theory.
We also conclude that one D7-brane is located between each vacua on $\mathds{S}^1$ at $\rho=0$. For large $N$, the distance between the branes is small and a 4D observer who measures low-energy interactions on $R^{1,3}$ sees a stack of $N$ D7-branes and $SU(N)$ gauge theory.

In the gauge theory, the different vacua arise due to gaugino condensation and it is believed that domain walls that separate the vacua exist in the IR, but not in the UV. This is consistent with the background we have here as separate vacua sit at distinct values of $\psi$ on $\mathds{S^1}$ which has largest size at the IR boundary, and $\mathds{S}^1$ gets smaller and becomes of zero-size at the UV boundary.

\section{Discussion\label{concl}}

The gravity theory involves simple analytic expressions for all
components of the metric and the fluxes and reproduces key features of the gauge theory, its symmetries, renormalization group flow, and chiral symmetry breaking.
The gravity background has a number of interesting features. The geometry is compact and the internal space normal to the D7-branes is $\mathds{S}^1$ at the IR boundary and spacetime is 4D at the UV boundary, consistent with the symmetries of the gauge theory and the radius of $\mathds{S}^1$ is set by the nonperturbative scale of the gauge theory. The compactness also provides a setting with well-defined vacuum and accommodates obtaining small nonzero gravitational constant in 4D from a large one in 10D. The background geometry does not need to be glued to other backgrounds. The curvature of the compact space is small and the supergravity flow is smooth in the IR region where the gauge theory is strongly coupled and a dual gravity description is useful. The range of the radial direction on $\mathds{R}^1$ gets smaller for larger $g_sN$, consistent with the strength of gauge coupling running in the gauge theory. The scale of string tension for an observer in 4D is of the same order and smaller than the scale of KK and glueball masses, which is useful for studying mass spectra in the gauge theory using the gravity theory. The UV boundary provides a convenient setting for putting quarks and antiquarks and a UV cutoff to the gravity theory, with appropriate choice of $g_sN$, beyond which the gauge gravity description is appropriate.
The classical supergravity description accommodates a range of physically interesting 't Hooft coupling.

Our objective in this paper was to produce the most basic features of the pure gauge theory within a simple setting of the gravity theory. It will be interesting to study additional features, implications, and tests of the gravity theory and the proposed correspondence.

The warp factor on the 4D spacetime in the metric given by (\ref{metric-3}) is nonzero everywhere and increases with increasing $r$.
One measure of confinement is the area enclosed by a Wilson loop \cite{Wilson:1974sk}. It follows from the metric that the area bounded by a Wilson loop in 4D spacetime at some $r>r_s$ is minimized for a surface stretched and bent toward $r=r_s$, the warp factor is nonzero and finite there, and the background leads to confinement. The specifics of confinement are presented in a separate paper \cite{Hailu:2011conf}, since the result is quite interesting on its own and the crucial component of the supergravity background used is the metric.

Glueball mass spectra can be computed by studying metric and flux fluctuations on the explicit background.

Because the gravity theory contains crucial features of the nonsupersymmetric strong nuclear interactions, gauge coupling running, chiral symmetry breaking, and confinement, it might provide a setting for exploring nonperturbative phenomena in QCD.

The metric might also be useful for exploring a supergravity implementation of \cite{Randall:1999ee} by generating a hierarchy of scales between probe branes at (or near) the UV boundary as hidden branes and the wrapped D7-branes at the IR boundary as visible branes in a compact extra dimensional space stabilized by a balance between fluxes and torsion.

In addition, the structure of the metric is such that the warp factor can be made nearly flat for a wide range of radial supergravity flow when $g_s N$ is small and may be useful for modeling inflationary scenarios of the early universe.

We would also like to point out that the supersymmetry equations of motion together with a physically motivated flux and/or metric provide a powerful approach to constructing interesting supergravity backgrounds systematically.

The magnitude of the 't Hooft coupling varies along the radial supergravity flow. It is large near the IR boundary and becomes smaller as the theory flows toward the UV boundary. The value of the 't Hooft coupling at the UV boundary, $g_sN$, can be chosen to be small such that perturbative gauge theory is the appropriate description in the region close to it. Because the supergravity flow between the IR and the UV boundaries is smooth, it may be possible to find a common region where both the gauge theory description and the gravity description, perhaps supplemented by $\alpha'$ corrections, could be studied and compared for a direct test of the gauge/gravity duality.

Finally, for a gauge/gravity duality to be useful for studying glueball and hadron mass spectra, it seems to us that the scale of string tension measured by a 4D observer needs to be of the same order as the scale of KK masses. Previous gauge/gravity constructions have large gap between the scales with the scale of string tension being much bigger than the scale of KK masses. The scale of KK masses in the IR is of the same order and bigger than the scale of string tension here, which is an appropriate feature for studying mass spectra in the gauge theory using the gravity theory, and it was obtained on a compact background that has small curvature in the IR.
This was possible because the 't Hooft coupling is large in the IR independent of $\alpha'$, the value of the radial coordinate and the size of the compact space at the IR boundary are set by the nonperturbative scale of the gauge theory, and smaller curvature in the IR is obtained with a larger range in the radial direction which corresponds to smaller 't Hooft coupling at the UV boundary. From the gauge theory point of view, this is because it takes a flow along a wider range of scales to get form large 't Hooft coupling at the IR boundary to a small one at the UV boundary.
It will be interesting to investigate $\alpha'$ corrections to the supergravity description and their interpretations.
The results in this paper and in \cite{Hailu:2011conf} show that the classical supergravity description captures key features of the nonperturbative IR dynamics of the gauge theory.

\providecommand{\href}[2]{#2}
\end{document}